\documentclass[10pt,preprint]{emulateapj}
\usepackage{url}
\usepackage{color}
\usepackage{natbib}
\newcommand{\fig}[1]{Figure~\ref{#1}}
\newcommand{\tbl}[1]{Table~\ref{#1}}
\newcommand{\speed}[1]{#1 km~s${}^{-1}$}

\newcommand{\acc}[1]{#1 m~s${}^{-2}$}
\newcommand{\rsun}[1]{${#1}\,R_\odot$}

\shorttitle{Successive Oscillating Filaments Triggered by an Invisible  EUV Wave}
\shortauthors{Shen et al.}
\begin{document}
\title{A Chain of Winking (Oscillating) Filaments Triggered by an Invisible Extreme-ultraviolet Wave}
\author{Yuandeng Shen\altaffilmark{1,2,3}, Kiyoshi Ichimoto\altaffilmark{2}, Takako T. Ishii\altaffilmark{2}, Zhanjun Tian\altaffilmark{1}, Ruijuan Zhao\altaffilmark{1}, and Kazunari Shibata\altaffilmark{2}}
\altaffiltext{1}{Yunnan Observatories, Chinese Academy of Sciences, Kunming 650011, China; ydshen@ynao.ac.cn}
\altaffiltext{2}{Kwasan and Hida Observatories, Kyoto University, Yamashina-ku, Kyoto 607-8471, Japan}
\altaffiltext{3}{Key Laboratory of Dark Matter and Space Astronomy, Purple Mountain Observatory, Chinese Academy of Sciences, Nanjing 210008, China}

\begin{abstract}
Winking (oscillating) filaments have been observed for many years. However, observations of successive winking filaments in one event have not been reported yet. In this paper, we present the observations of a chain of winking filaments and a subsequent jet that are observed right after the X2.1 flare in AR11283. The event also produced an Extreme-ultraviolet (EUV) wave that has two components: upward dome-like wave (\speed{850}) and lateral surface wave (\speed{554}) which was very weak (or invisible) in imaging observations. By analyzing the temporal and spatial relationships between the oscillating filaments and the EUV waves, we propose that all the winking filaments and the jet were triggered by the weak (or invisible) lateral surface EUV wave. The oscillation of the filaments last for two or three cycles, and their periods, Doppler velocity amplitudes, and damping times are 11 -- 22 minutes, \speed{6 -- 14}, and 25 -- 60 minutes, respectively. We further estimate the radial component magnetic field and the maximum kinetic energy of the filaments, and they are 5 -- 10 Gauss and $\sim 10^{19} \, {\rm J}$, respectively. The estimated maximum kinetic energy is comparable to the minimum energy of ordinary EUV waves, suggesting that EUV waves can efficiently launch filament oscillations on their path. Based on our analysis results, we conclude that the EUV wave is a good agent for triggering and connecting successive but separated solar activities in the solar atmosphere, and it is also important for producing solar sympathetic eruptions.
\end{abstract}

\keywords{Sun: activity --- Sun: flares --- Sun: filaments, prominences --- Sun: oscillations --- Sun: magnetic fields}

\section{Introduction}
A solar filament or prominence is made up of cool plasma, often in a loop shape, extending outward from the Sun's surface into the extremely hot corona. It is typically one hundred times cooler and denser than the surrounding coronal plasma. The filament and prominence are the same entities with the former against the solar disk and latter over the limb, and they appear darker and brighter than their surrounding backgrounds, respectively. Hereafter, we use the term ``filament'' throughout the paper. Previous studies indicate that magnetic field always plays a key role in the whole lifetime of a filament \citep{kipp57,kupe74,tand95}, and typically, whose eruption is often accompanying with  flares and coronal mass ejections (CMEs) \citep[e.g.,][]{shen11b,shen12d,yan12a,yan12b,bi12,bi13a,chen13}. So far, many questions about filaments and their eruptions remain unresolved, and the investigation of those questions, such as how and why are they formed, as well as their stability, eruption mechanisms, and potential impact on physical condition of the terrestrial space, have attracted a lot of attentions of solar physicists in recent decades.

The phenomenon of filament oscillation is important due to the possible application in filament seismology that is a new and promising technique for estimating physical parameters of filaments and their surrounding coronal plasma. As early as in the 1930s, solar physicists have noted the existence of periodic oscillating motion in filaments \citep{dyso30,newt35}. Since then, a number of observational and theoretical studies have been performed for understanding the physical nature of the filament oscillation \citep[e.g.,][]{dods49,more60,hyde66,rams66}. According to the velocity amplitude, filament oscillations can be classified into two categories, i.e., small and large amplitude oscillations. The former is quite common in filaments, which is often observed in a restricted volume of the filament body and is not related to flare activities. Such kind of oscillation usually has small velocity amplitudes of \speed{2 -- 3} and periods of 10 -- 80 min \citep{arre12,hill13}. The report of large amplitude oscillation is relatively scarce. It is often triggered by large-scale magnetohydrodynamics (MHD) waves or shocks (e.g., Moreton waves \citep{more60}, or ``EIT'' waves \citep{okam04}) in association with remote flaring activities \citep{atha61,hyde66,rams66,eto02,okam04,asai12,shen12b,shen12c,jack13}, or near-by micro-flares and jets \citep{jing03,jing06,vrsn07,li12}. The velocity amplitude of large amplitude oscillation is often of tens of \speed{}, while the period is typically of 6 -- 150 minutes \citep{trip09}. Particularly, long and ultra-long period (8 -- 27 hours) of oscillations are also observed in filaments \citep{foul04,foul09}. In a special case, oscillation was observed before and during the slow-rising, pre-eruption phase of an erupting filament \citep{isob06,isob07,pint08,chen08}. The authors proposed that the fast magnetic reconnection that changes the equilibrium of the supporting magnetic system could be the possible trigger mechanism, and such kind of activity is important for diagnosing and forecasting filament eruptions.

On the other hand, according to the oscillating direction relative to the filament's main axis, large amplitude oscillation could also be divided into longitudinal \citep[e.g.,][]{zirk98,jing03,vrsn07,zhan12,zhan13,luna12a,luna12b}, horizontal \citep[e.g.,][]{klec69,hers11,shen12b}, and vertical oscillations \citep[e.g.,][]{hyde66,eto02,okam04}. For vertical (horizontal) large amplitude oscillation of a filament close to (far away from) the disk center, one can observe the alternative appearance and disappearance of the filament body in H$\alpha$ center and line-wings due to the oscillating movement of the filament, which causes the Doppler shift from the H$\alpha$ center to the red wing and then the blue wing. This process repeats periodically and for this reason suck kind of oscillating filament was dubbed ``winking filament'' in history. \cite{rams66} and \cite{hyde66} studied 11 winking filaments and found that the oscillation periods are not related to the filament dimensions and the power and distance of the disturbing flare, namely, filaments always oscillate with their characteristic periods. \cite{hers11} also confirmed such a result, and they further proposed that large amplitude filament oscillations are actually a collection of separate but interacting fine threads.

As mentioned above, there are several possible agents that can trigger large amplitude filament oscillations. For winking filaments, they are often triggered by large-scale MHD waves or shocks in association with remote flare/CME activities. \cite{eto02} studied a winking filament that is associated with both a Moreton and an EUV waves. They found that the speed of the Moreton wave is approximately three times of the EUV wave, and the arriving of the inferred Moreton wave front is consistent with the start time of the filament oscillation, which suggests that the trigger of the filament oscillation should be the Moreton wave. In a statistical study performed by \cite{okam04}, the authors found that some winking filaments are triggered by EUV waves rather than Moreton waves, which lead the conclusion that EUV waves could be another possible trigger of winking filaments. In some events, one can not detect notable Moreton waves in chromosphere, even though the EUV wave is remarkable in coronal observations. Therefore, the observation of winking filaments can be used to diagnose the existence and the properties of invisible (weak) Moreton waves \citep[e.g.,][]{rams66,eto02,okam04}. For example, based on observations of winking filaments, \cite{eto02} found that the Moreton wave was inconsistent with the corresponding coronal EUV wave. \cite{gilb08} found that there is a delay in the filament rebound accompanying the initial compression of the wave front, which suggests the filament is sensitive to the width of the wavefront. Their study also indicated that the wavefront topology should be slightly forward-inclined with respect to the surface, confirming the speculation of \cite{rams66} and the theoretical prediction in the numerical model of Moreton wave \citep{uchi68}, where the Moreton wave is explained as the intersection line between an expanding coronal wave front surface and the dense chromosphere. A few recent studies using high-resolution observations also confirm the forward-inclined topology of Moreton waves \citep[e.g.,][]{liu12,liu13}.

The study of winking (oscillating) filaments opens a new discipline called filament seismology. In general, the filament seismology technique is based on the application of theoretical knowledge and inversion techniques using the observed periods, damping times, and flow speeds to estimate the physical parameters of the filaments and the surrounding coronal plasma. With the application of \cite{kipp57} model of filaments into winking filaments reported by \cite{rams66}, \cite{hyde66} obtained the radial components of the filament magnetic fields ($B_{\rm r}$) are ranging from 2 to 30 Gauss, while the effective coefficients of viscosity ($\eta$) is from $4 \times 10^{-10}$ to $1.6 \times 10^{-9}$ poise. By assuming the temperature ($T = 10^{6}$ K) and electron number density ($n_{\rm e} = 10^{9}$ ${\rm cm}^{-3}$) in the lower corona, the author further derived the coronal magnetic fields ($B_{\rm e}$) surrounding the filaments, which is about 0.09 -- 0.18 Gauss. The inferred magnetic fields of filaments are in agreement with those obtained from direct measurements \citep[e.g.,][]{ziri61,lee65} and from the analysis of the polarization of H$\alpha$ and \ion{He}{1} $D_{\rm 3}$ lines \citep[e.g.,][]{hyde65,warw65}. The study of \cite{hyde66} demonstrates that filament seismology is a reasonable method for deriving various physical parameters of filaments. In addition, filament oscillations are also important for diagnosing the stability and eruption mechanism of filaments \citep{isob07,pint08}. For more background knowledge on filament oscillation and the application in filament seismology, we refer to several recent reviews \citep{oliv02,trip09,arre12}.

In this paper, we present an interesting observational study of a chain of winking filaments that was in association with a {\it GOES} X2.1 flare in NOAA active region AR11283 (N13W18) on September 06, 2011. The flare produced with a remarkable EUV wave propagating mainly in the northwest direction, which not only triggered the oscillation of three filaments in the northwest of AR11283, it also launched the oscillation of a long filament and the occurrence of a small jet in the eastern hemisphere where the wave signature is very weak or even invisible. Thanks to the excellent observations obtained by the Solar Magnetic Activity Research Telescope \citep[SMART;][]{ueno04}, and the {\it Solar Dynamics Observatory} \citep[{\it SDO};][]{pesn12}, we are able to study the winking filaments and the EUV wave in great detail. In Section 2, observations and measuring methods are described briefly. Section 3 is the main analysis results. Conclusions and Discussions are given in Section 4.

\section{Data Sets \& Methods}
\subsection{Data \& Instruments}
In this study, we mainly use the observations taken by the SMART and {\it SDO} instruments. The SMART is a ground-based telescope at Hida Observatory of Kyoto University, Japan, which is designed to observe the dynamic chromosphere and the photospheric magnetic field. The SMART consists of four different telescopes with apertures of 20 to 25 cm, including T1 for H$\alpha$ full-disk observations, T2 for vector magnetic field measurements, T3 for H$\alpha$/continum high speed observations \citep{ishi13}, and T4 for measurements of photospheric vector magnetic field \citep{naga14}. We use the full-disk H$\alpha$ images taken by T1, which has seven wavelength channels close to the H$\alpha$ line-center, i.e., H$\alpha$ center, $\pm 0.5$ \AA, $\pm 0.8$ \AA, and $\pm 1.2$ \AA. The telescope takes images with the seven channels near simultaneously ($< 1$ min), and the field-of-view (FOV) and pixel resolution are $2300\arcsec \times 2300\arcsec$ and $0\arcsec.6$, respectively. The cadence of the images is 2 minutes. The advantage of this telescope is that it can monitor any solar activity on the disk above the spatial resolution. Especially, the H$\alpha$ off-bands observations are very suitable for detecting chromospheric dynamic activities such as the activation, oscillation, and eruption of filaments and other chromospheric structures. 

The space-borne {\it SDO} carries three instruments, namely, the Atmospheric Imaging Assembly \citep[AIA;][]{leme12}, the Helioseismic and Magnetic Imager \citep[HMI;][]{scho12}, and the EUV Variability Experiment \citep[EVE;][]{wood12}. The SDO was launched on February 11, 2011 for studying the causes of solar variability and its impacts on Earth. In this study, we use the imaging observations taken by AIA and the line-of-sight (LOS) magnetograms provided by HMI. AIA has high time cadences up to 12 s and high pixel resolution of $0\arcsec.6$. It captures images of the Sun's atmosphere out to \rsun{1.3} in seven EUV and three UV-visible wavelengths. The cadence of HMI magnetograms is 45 s, while the measurement precision is 10 Gauss. All images used in this paper are differentially rotated to a reference time (22:20:00 UT), and the solar north is up, west to the right.

\subsection{The Cloud Model}
Since the direct measuring of filament parameters is difficult, solar physicists have designed alternative methods to derive the parameters indirectly. For many years, a widely used method is the so-called ``cloud model'' that was proposed by \cite{beck64}. For example, \cite{mori03,mori10} used the model to derive the three-dimensional velocity of erupting filaments using H$\alpha$ observations taken by the Hida Flare Monitoring Telescope \citep[FMT;][]{kuro95}.

The basic idea of cloud model is that a filament can be considered as a cloud lying between observer and the background chromosphere. In the cloud model, it assumes that the filament is fully separated from the underlying chromosphere, and the background intensity is the same below the filament and the surrounding atmosphere, and the source function, radial velocity, Doppler width and the absorption coefficient are constant along the LOS. Generally, an observed intensity line profile is the function of several parameters describing the three-dimensional solar atmosphere, and it can be described by the radiative transfer equation
\begin{equation}
I (\Delta \lambda) = I_{\rm 0}(\Delta \lambda) \, e^{-\tau (\Delta \lambda)} + \int_{\rm 0}^{\tau (\Delta \lambda)} S_{\rm t} \, e^{-t (\Delta \lambda)} \, dt,
\end{equation}
where $\Delta \lambda$ represents the wavelength difference between the observation wavelength and the H$\alpha$ line-center, $I (\Delta \lambda)$ is the observed intensity, $I_{0}(\Delta \lambda)$ is the reference background intensity, $\tau (\Delta \lambda)$ is the optical thickness, $S$ is the source function which is a function of optical depth along the cloud, $t$ is an intermediate variable related to $\tau (\Delta \lambda)$. With the assumptions made in the cloud model, the radiative transfer equation describing the observed intensity line profile could be reduced to 
\begin{equation}
I(\Delta \lambda) = I_{\rm 0}(\Delta \lambda) \, e^{-\tau(\Delta \lambda)} + S \, (1 - e^{-\tau(\Delta \lambda)}),
\end{equation}
and can be rewritten as the form of contrast profile
\begin{equation}
C(\Delta \lambda) = \frac{I(\Delta \lambda) - I_{\rm 0}(\Delta \lambda)}{I_{\rm 0}(\Delta \lambda)} = (\frac{S}{I_{\rm 0}(\Delta \lambda)} - 1) \, (1 - e^{- \tau (\Delta \lambda)}),
\label{const}
\end{equation}
where $\tau (\Delta \lambda)$ is the optical depth, which has a Gaussian wavelength dependence form and can be written as follows
\begin{equation}
\tau (\Delta \lambda) = \tau _{\rm 0} \, e^{- (\frac{\Delta \lambda - \Delta \lambda _{\rm I}}{\Delta \lambda _{\rm D}})^2},
\end{equation}
where $\tau _{\rm 0}$ is the line-center optical thickness, $\Delta \lambda _{\rm I} = \frac{\lambda_{\rm 0}v}{c}$ is the Doppler shift with $\lambda_{\rm 0}$ being the line-center wavelength, $c$ the speed of light, and $\Delta \lambda _{\rm D}$ the Doppler width which depends on temperature $T$ and micro-turbulent velocity $\xi_{\rm t}$, i.e., 
\begin{equation}
\Delta \lambda_{\rm D} = \frac{\lambda_{\rm 0}}{\rm c} \sqrt{\xi_{\rm t}^{2} + \frac{2kT}{m}},
\end{equation}
where $m$ is the atom rest mass.

As one can see, there are four unknown parameters in equation \ref{const}, namely the source function $S$, the Doppler width $\Delta \lambda_{\rm D}$, the optical thickness $\tau _{\rm 0}$, and the LOS velocity $v$. In the calculation, all these parameters are assumed to be constant along LOS in the cloud. Finally, given at least four contrast values, one can derive these unknown parameters. Fortunately, the SMART provides seven wavelengths H$\alpha$ observations simultaneously, which is sufficient for us to derive the Doppler velocity of the oscillating filaments using the cloud model. We will show the inversion results in the next Section.

\begin{figure*}[thbp]
\epsscale{1}
\plotone{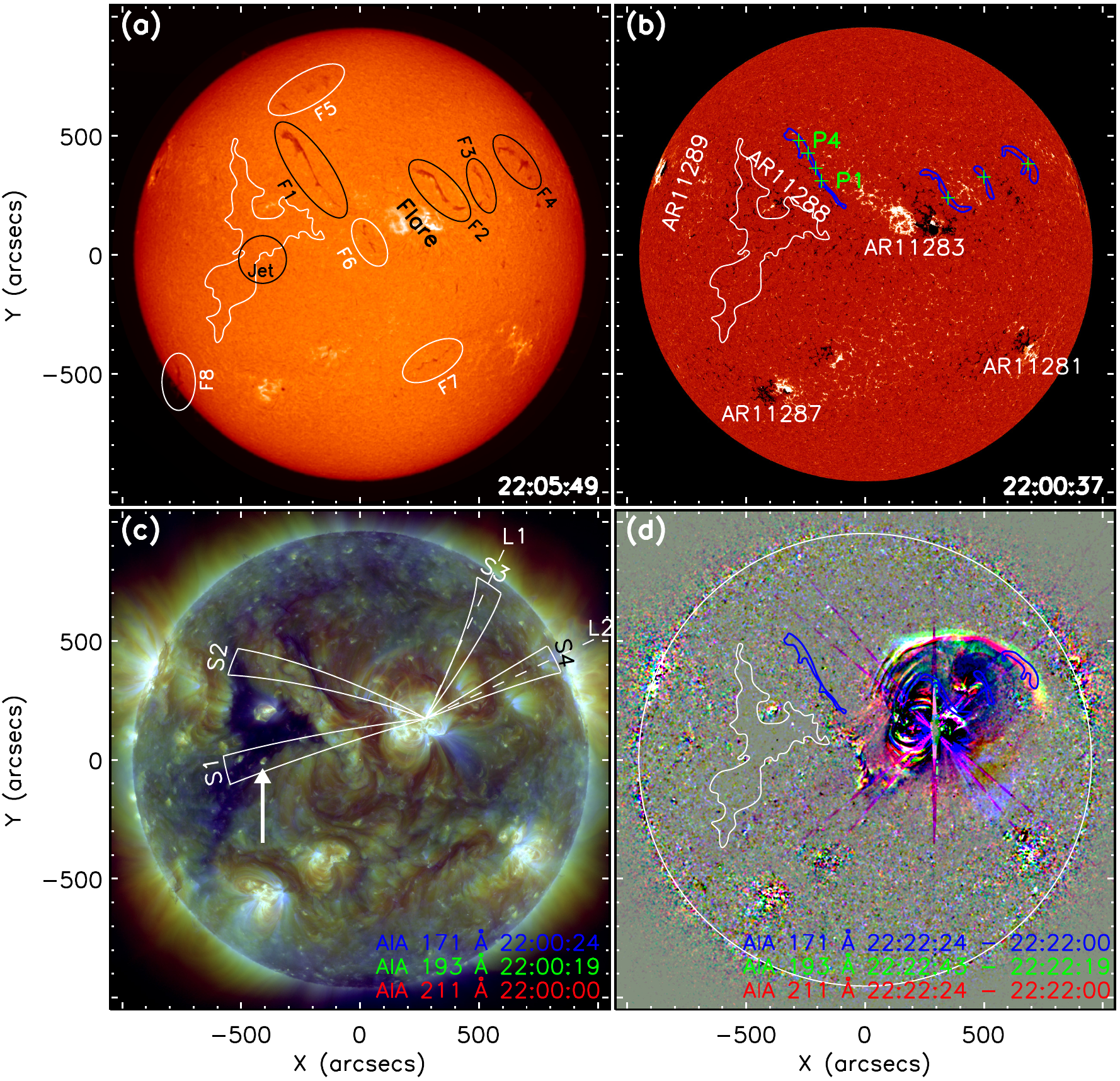}
\caption{An overview of the event before the X2.1 flare. Panel (a) is a SMART H$\alpha$ center image, in which the oscillating filaments are highlighted with black ellipses (F1 -- F4), while the white ellipses indicate the stable ones (F5 -- F8). The black circle indicates the jet, and the white contour marks the coronal hole boundary. Panel (b) is an HMI LOS magnetogram overlaid with contours of the coronal hole and the oscillating filaments. The green crosses in panel (b) are the position where Doppler velocities are measured. Panel (c) and (d) are raw and running-difference tri-color images, in which the blue, green, and red color represent the AIA 171, 193, and 211 \AA\ channels, respectively. The sectors (S1 -- S4) and the dashed lines (L1 and L2) in panel (c) are used to obtain time-distance plots. The white arrow points to the bright point at the jet-base. The FOV of each frame is $2100\arcsec \times 2100\arcsec$. Animation of this figure is available in the online journal.
\label{fig1}}
\end{figure*}

\section{Results}
\subsection{Overview of the Event}
The NOAA active region AR11283 became very energetic after evolving into a complex $\beta\delta$ active region from a simple $\alpha$ type sunspot. Besides a number of B and C flares, it produced two X and five M flares from September 06 to 10, 2011. On September 06, the active region produced the most powerful X2.1 flare during its lifetime on the visible disk. According to the record of {\it GOES}, the start, peak, and end times of this flare are 22:12, 22:20, and 22:24 UT, respectively. Based on high temporal and spatial resolution observations taken by {\it SDO} and magnetic extrapolation techniques, several studies about this flare and the associated phenomena have been published \citep{jian13,jian14a,jian14b,feng13,dai13,roma13,verw13,nitt13,ruan14}. The authors investigated the energy accumulation and triggering mechanism of the flare, the associated CME, the three-dimensional magnetic field above the active region, and the oscillation coronal loops caused by the flare. We find that this flare was in association with a global EUV wave which resulted in the oscillation of four remote filaments, as well as a small jet close to the boundary of a coronal hole in the eastern hemisphere. A detailed investigation of these associated phenomena is useful for diagnosing the physical properties of both the filaments and the EUV wave, and these activities were not analyzed in the previous articles. Therefore, we take this as the main subject of this paper.

On September 06, AR11283 was close to the disk center (N14W18), and it appeared as a bright patch on H$\alpha$ line-center images before the flare. In addition, there are eight filaments, a large trans-equatorial coronal hole, and four other active regions (ARs: 11281, 11287, 11288, and 11289) on the visible solar disk (see \fig{fig1}). In the H$\alpha$ line-wing observations, the four filaments circled by black ellipses (F1 -- F4) were disturbed and underwent obvious oscillations (seen as winking filaments). However, the other four (F5 -- F8) did not show any discernible activities. It is noted that the activation of the filament oscillations were in a proper time order depending on their distance to the flare. This suggests that the oscillation of the filaments were possibly launched by a common propagating disturbance from the flaring region. Besides the chain of filament oscillations, a subsequent small jet is also observed near the southern boundary of the coronal hole (see the black circle in \fig{fig1} (a)). In the EUV image shown in \fig{fig1}(c), the low temperature coronal hole manifested as a dark triangle region, and it is noted that there is a small bright point at the location of the jet, which is thought to be a small closed magnetic field system that provides the necessary condition for producing the jet. \fig{fig1}(d) is a tri-color image created from AIA 171, 193, and 211 \AA\ running-difference images at about 22:22:24 UT, in which the blue, green, and red colors represent the three different channels, respectively. We can see that the bright front of the EUV wave was mainly propagating in the northwest direction, which sweeps over filaments F2 -- F4. It seems that the EUV wave did not pass through the locations of F1 and the coronal jet in the east hemisphere. In the southwest direction, the signature of the EUV wave was very weak (see animation 2). However, when the wave passing through AR11281, diffraction effect the EUV wave, which was first reported by \cite{shen13a}, can be identified by seeing the time sequence AIA 211 \AA\ running-ratio difference images, which suggests that the physical nature of this EUV wave should be a fast-mode magnetosonic wave in nature.

According to previous studies, winking filaments are often trigged by either chromospheric Moreton waves or coronal EUV waves \citep{eto02,okam04}. In the present case, we do not detect any significant signature of Moreton wave in H$\alpha$ observations. On the other hand, since the EUV wave mainly propagated in the northwest of AR11283, it is hard to understand the trigger mechanisms of the F1's oscillation and the occurrence of the small coronal jet.

\subsection{The EUV wave}
To analyze the kinematics of the EUV wave on September 06, we generate time-distance plots along four sectors S1 -- S4 as shown in \fig{fig1}(c). The width of each sectors is 10${}^\circ$, and each one is composed by two great circles, whose crossing point is the flare kernel. From each image, an one-dimensional intensity profile as a function of distance within a sector can be obtained by averaging the intensities across the sector, in annuli of increasing radii with 0.05${}^\circ$ that corresponds a distance of 608 km. Composing the obtained one-dimensional profiles over time yields a two-dimensional time-distance plot along the sector, in which the propagating EUV wave shows up as an incline ridge whose slope represents the apparent speed along the solar surface. This method can effectively minimize the spherical projection effect, and it has been widely used for diagnosing the kinematics of EUV waves in previous studies.

\begin{figure*}[thbp]
\epsscale{0.95}
\plotone{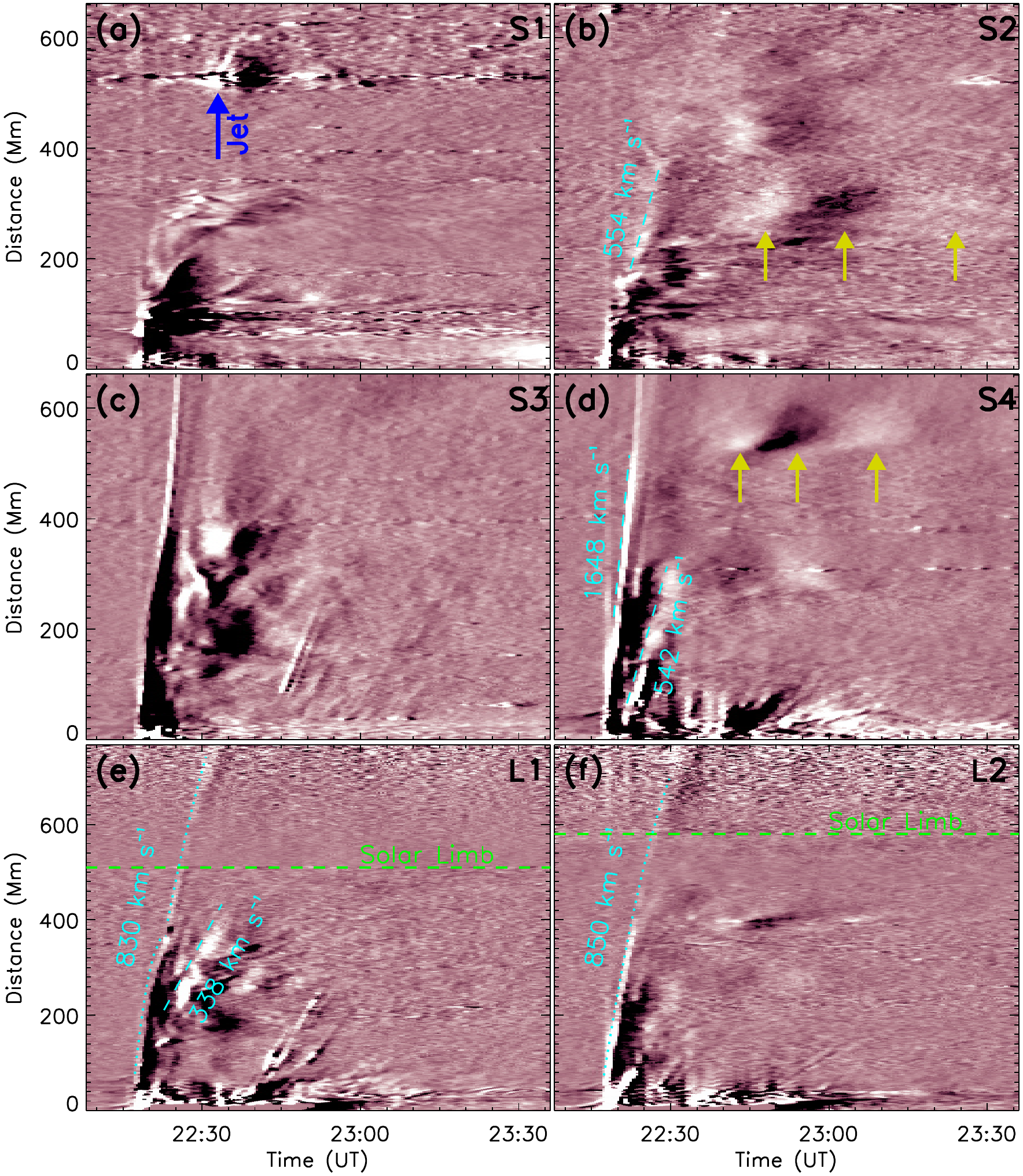}
\caption{Time-distance plots show the detailed kinematics of the EUV wave along the four sectors and the two dashed lines shown in \fig{fig1}(c). Panel (a) -- (d) are obtained from 211 \AA\ running-ratio images along sectors S1 -- S4, while panels (e) and (f) are along L1 and L2, respectively. The dashed (dotted) lines are linear (quadratic) fitting to the wave stripes, the arrow in panel (a) points to the jet, while those in panels (b) and (d) indicate the intensity fluctuation patches after the passage of the EUV wave. The green dashed lines in panels (e) and (f) indicate the limb of the solar disk.
\label{fig2}}
\end{figure*}

The time-distance plots shown in \fig{fig2} are made from AIA 211 \AA\ running-ratio images, which are generated by firstly subtracting each image from the one taken at 24 s before, and then dividing the previous image, i.e., $I_{\rm ratio} = \frac{I_{\rm t} - I_{\rm t-24}}{I_{\rm t-24}}$. Such images can protrude the fast changing and propagating features such as EUV waves. Consistent with the imaging observation, in the time-distance plot along S1 that passes through the jet-base, we can not observe any significant wave signature but the the jet, which is indicated by a red arrow in panel (a). S2 passes through the middle section of F1, along which the wave signature is very weak. Making a linear fit to the stripe yields the wave speed, which is about \speed{554}. After the passing of the EUV wave, we can observe a region of intensity fluctuation behind the wave front, which is indicated by the green arrows in panel (b). We propose that the intensity fluctuation is probably caused by the oscillation of F1, even though it can not be identified in imaging observations. S3 is placed on a quiet-sun region where the wave is the most remarkable. In the time-distance plot along S3, we find that the wave suddenly changed its propagation direction at a distance of about 400 Mm from the flare kernel, where one can identify a small magnetic structure over there. Therefore, the changing of the propagation direction could be interpreted as the refraction of the EUV wave caused by the small magnetic structure on the path, as what has been reported in \cite{shen12b}. Such a phenomenon can be taken as a direct evidence supporting the scenario that EUV waves are fast-mode magnetosonic waves in nature. Another interesting feature is the thin parallel stripes observed during 22:40 UT to 22:55 UT, which could probably be interpreted as the so-called quasi-periodic fast propagating magnetosonic wave trains associated with flare pulsations \citep[e.g.,][]{liu11,shen12a,shen13b}. S4 passes through the oscillating filaments of F2 -- F4. In the time-distance plot, a fast and a slow EUV waves can be observed simultaneously, and their speeds are about \speed{1648} and \speed{542}, respectively. The speed of the fast wave is about three time of the slower one, in agreement with the previous observational and numerical studies \citep{shen12c,shen13a,chen02,chen11,xue13}, where the fast wave is interpreted as fast-mode MHD wave or shock, while the slow one is caused by the field line reconfiguration caused by the associated CME. Similarly, intensity fluctuation is also observed after the passing of the fast EUV wave (see the green arrows in \fig{fig2}(d)). 

Since sectors S3 and S4 are toward the limb, the measurement of the wave speed along the two sectors are less credible than that close to the disk center. Therefore, we alternatively measure the projection speed on the plane of the sky along two straight lines (L1 and L2) as shown in \fig{fig1} (c). The corresponding time-distance plots along L1 and L2 are shown in \fig{fig2} (e) and (f), respectively. In the time-distance plots, the wave stripe is not linearly, and it can be fitted with a quadratic function. One can see that the EUV wave propagates fast at the initial stage, and then it  decelerates to a moderate speed. From the time-distance plot along L2, the speeds of the initial and final stages are about \speed{1083 and 678}, respectively, while the deceleration is about \acc{-224}. The average speed of the fast EUV wave is about \speed{850}, and that is about \speed{338} for the following slow EUV wave. It can be seen in \fig{fig2} (e) and (f) and animation 2, the propagation of the fast EUV wave did not show any changes when it passes over the solar disk limb (see the green dashed line). As shown in \cite{vero10}, an EUV wave usually has a dome shape in the upward direction, and on the surface one can observe another EUV wave that is thought to be driven by the associated CME at the initial stage, and then it will propagates freely on the surface \citep{vero10}. Therefore, the wave speed measures along L1 and L2 should be the projection upward speed of the EUV wave (dome part), while that along S2 should be the speed of the surface wave. Therefore, the speed difference in different directions could be understood from the topology of EUV waves.

\begin{figure*}[thbp]
\epsscale{0.95}
\plotone{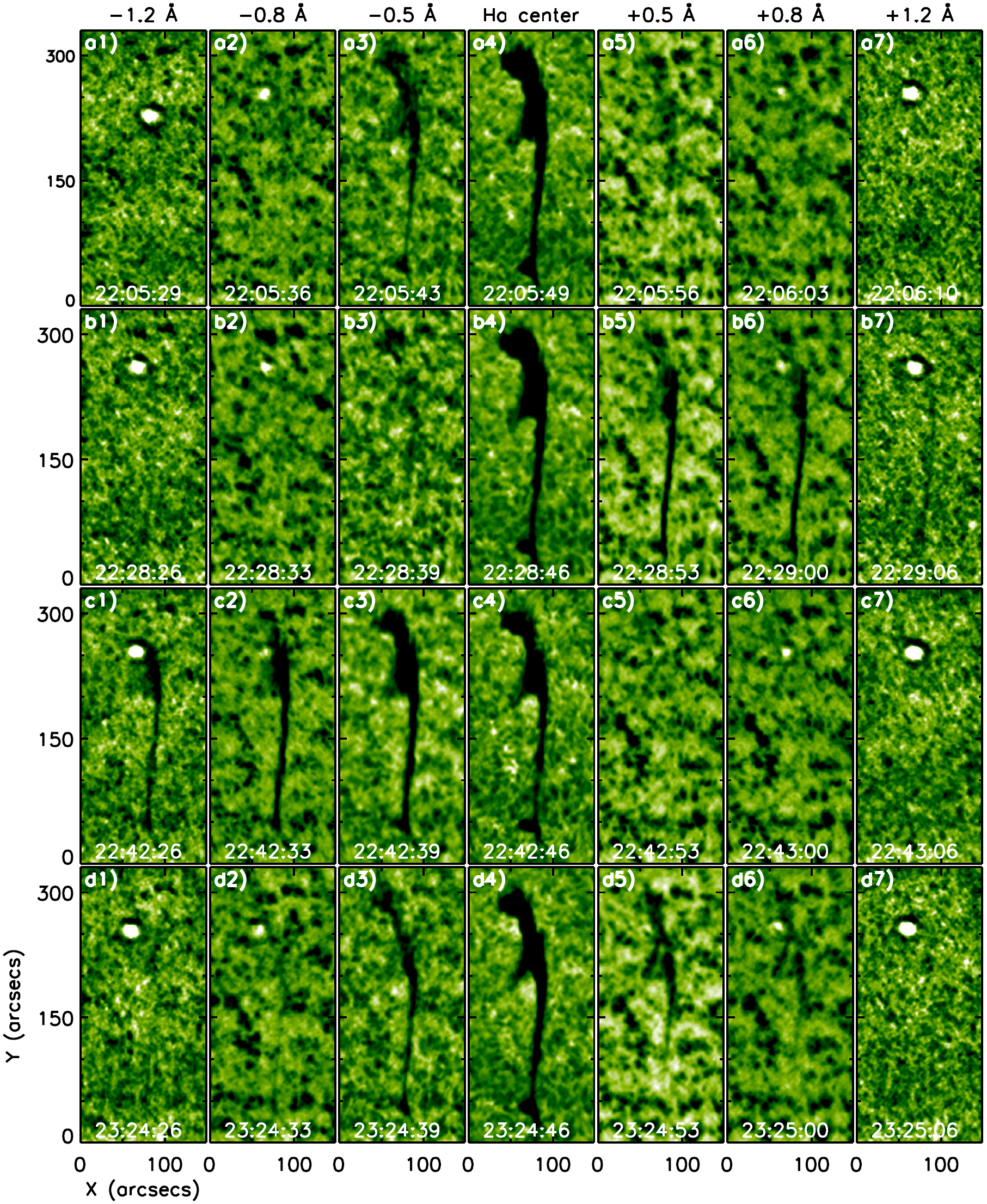}
\caption{Time sequence of SMART H$\alpha$ images show the oscillating F1. From left to right, they are -1.2, -0.8, -0.5, 0.0, +0.5, +0.8, and +1.2 \AA\ images shifting from the H$\alpha$ line-center (6562.8 \AA). For each row, the images are taken at a near simultaneous time ($\Delta t < 1$ minute). It should be noted that the images are rotated clockwise for 30$\degr$. The FOV for each frame is $150\arcsec \times 330\arcsec$. The bright point at the north end of F1 in $\pm 1.2$ \AA\ images is possibly caused by the SMART instrument.
\label{fig3}}
\end{figure*}

\begin{figure*}[thbp]
\epsscale{0.95}
\plotone{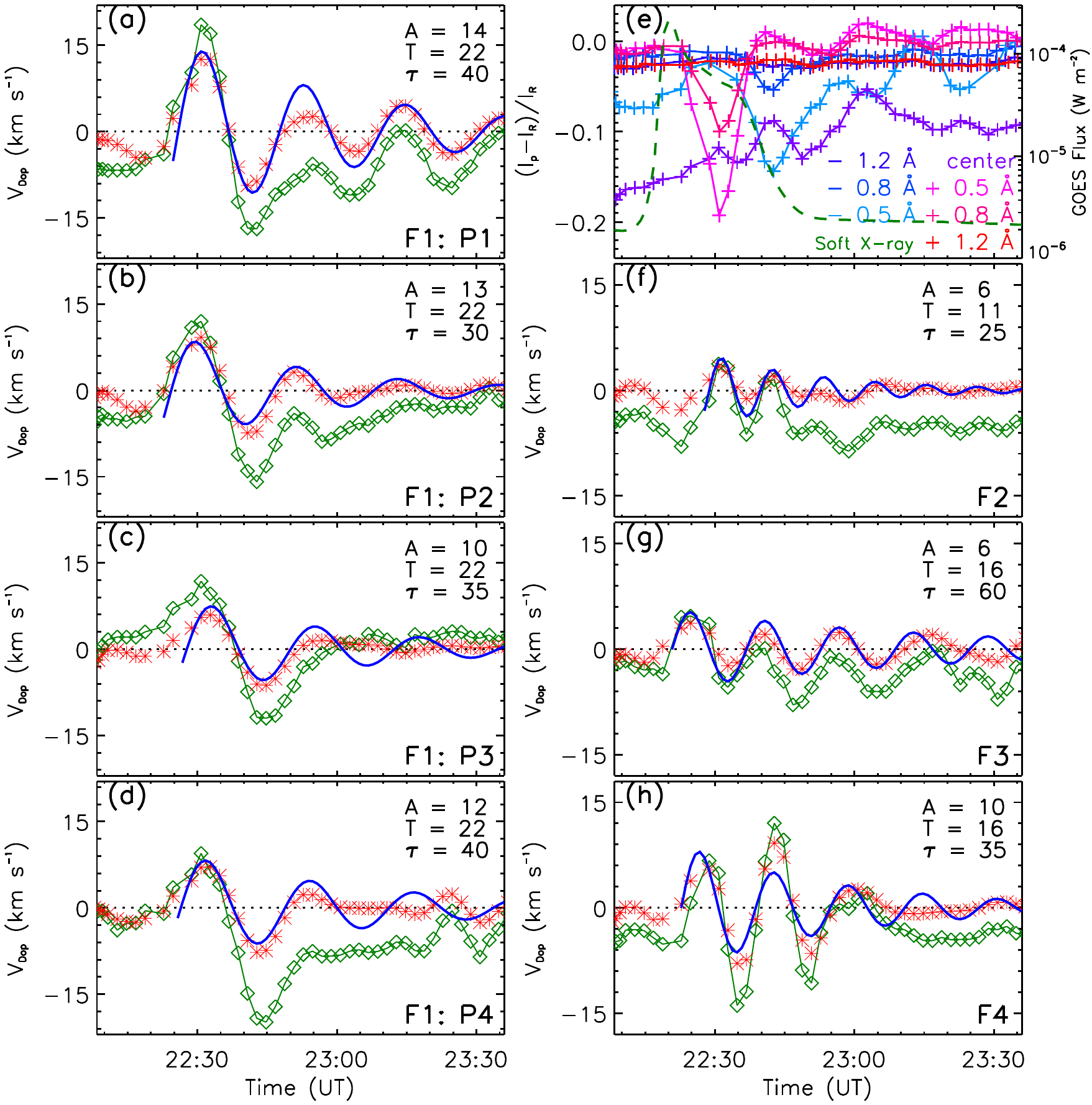}
\caption{The measurement of the Doppler velocity of the oscillating filaments. The Doppler velocities at the four points along F1 are shown in panels (a) --(d), while those for F2 -- F3 are plotted in panels (f) -- (h), respectively. In each panel, the green curve with diamond is the derived Doppler velocity, while the red curve with asterisk is the detrended velocity obtained by subtracting the smoothed velocity using a 10 minutes boxcar. The blue curve is a fit to the detrended data points in a form of $F(t) = A \exp(-\frac{t}{\tau}) \sin(\omega t+\phi)$. The velocity amplitude ($A$), oscillation period ($T$), and damping time ($\tau$) are also plotted in the figure. Panel (e) is the plot of the intensity contrasts of the seven H$\alpha$ wavelengths at P1, in which the green dashed curve is the {\it GOES} 1 -- 8 \AA\ soft X-ray flux.
\label{fig4}}
\end{figure*}

\subsection{The Filament Oscillations}
In the present case, the four oscillating filaments showed similar oscillation behaviors. They all underwent a damping oscillating process which lasted for about two or three cycles, after which the oscillating filaments regained their equilibrium rather than eruption. A reason for this is possibly that the disturbance to the filaments is too weak, or the filaments were strongly confined by the overlying coronal magnetic fields \citep{jian14b}. As an example, the time sequence of H$\alpha$ images are  shown in \fig{fig3} to illustrate the oscillation process of F1. In the figure, the time sequence of H$\alpha$ line-center images are shown in the middle column, while the left (right) columns are the blue (red) wing images, which show the upward (downward) motions of the filament along the LOS. 

As shown in \fig{fig3}, the shape and the intensity of the filament body showed little difference during the entire oscillation period. In addition, there is no obvious horizontal oscillatory motions could be detected. These results indicate that the oscillation amplitude should be a small one, and the oscillation direction is mainly along the LOS. The first row show the filament right before the start of the oscillation. It can be seen that the filament in the H$\alpha$ line-center composes with a long spine and two barbs on the left side. At the same time, the entire filament structure can also be observed in the -0.5 \AA\ image. Since this filament located in the east hemisphere, the appearance of the filament on the blue wing is probably caused by the solar rotation, or an offset of filter wavelength due to a setting error. At around 22:28 UT, the filament appeared on the red-wing observations, suggesting that the filament was firstly pushed down abruptly. It should be noted that the intensity of the filament in the red-wing observations decreases with the increasing of the wavelength (see \fig{fig3} b5 to b7), which suggests the shifting of the wavelength of the H$\alpha$ line. About 14 minutes later, the filament changed its oscillation to upward motion, and it showed up in the blue-wing images (see \fig{fig3} c1 -- c3). At about 23:24 UT, the filament changed its oscillation again to downward motion, i.e., red shift. Comparing with the red-wing observations at 22:28 UT, this time the intensity of the filament decreased a lot due to the decreasing of the oscillation amplitude. These observations indicate that the filament experienced an up-and-down and decaying oscillation process. The oscillation of the filament indicates that the initiation of the oscillation is probably caused by the compression of the global EUV wave, and the resultant force of gravity and magnetic tension could be the restoring force of the oscillation.

The detailed analysis of the four oscillating filaments are shown in \fig{fig4}, in which panel (e) shows the intensity contrasts of the seven H$\alpha$ wavelengths at position P1 on F1 (see \fig{fig1}(b)). The {\it GOES} 1 -- 8 \AA\ soft X-ray flux is also overlapped in the panel to show the temporal relationship between the flare and the filament oscillation. The other panels show the Doppler velocities of the oscillating filaments (F1 -- F4), which are calculated from the intensity contrasts using the Beckers' cloud model \citep{beck64}. This method has been extensively applied to derive physical parameters of cloud-like structures in the solar atmosphere. A brief description of this method is given in Section 2.2, more details of the cloud model have been documented in many articles \citep[e.g.,][and the references therein]{tzio07}.

\begin{figure*}[thbp]
\epsscale{0.95}
\plotone{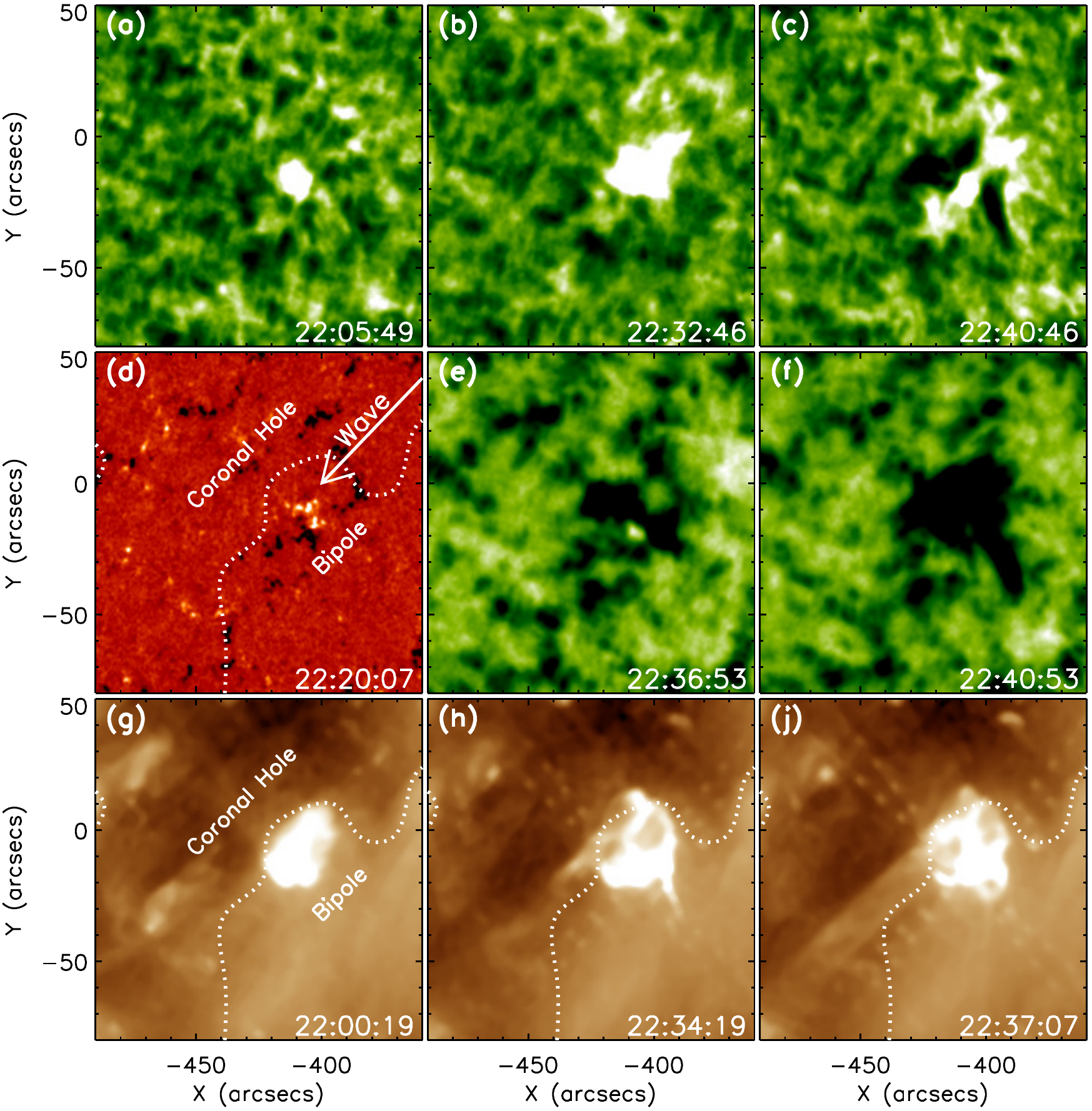}
\caption{SMART H$\alpha$ line-center ((a) -- (c)) and +0.5 \AA\ ((e) -- (f)) and AIA 193 \AA\ ((g) -- (j)) images show the ejection of the small jet at the southern boundary of the coronal hole. Panel (d) is an HMI LOS magnetogram showing the magnetic field configuration. The boundary of the coronal hole is overlaid on the magnetogram and 193 \AA\ images as a dotted curve, while the white arrow indicates the propagation direction of the EUV wave. The FOV for each frame is $220\arcsec \times 220\arcsec$.
\label{fig5}}
\end{figure*}

The variation of the intensity contrasts at P1 clearly show the oscillation behavior of the filament. On the red-wing, the highest contrast appeared at about 22:31 UT, while that is about 22:42 UT on the blue-wing. This result indicates that the oscillation half-period is about 11 minutes, and therefore the oscillation period of the filament should be about 22 minutes. The intensity contrasts indicate that the oscillation has a relative larger amplitude at the first, then it quickly decreases to a moderate amplitude. Such an evolution pattern implies that the filament was firstly pushed down due to a strong propagating pulse-like disturbance, for example, an EUV wave, and after the passage of the pulse, the filament starts to oscillate with its own characteristic period.

Panels (a) -- (d) show the Doppler velocities at the four positions along the main axis of F1 (see the red cross symbols in  \fig{fig1}(b)). The data points show an obvious oscillation behavior (green diamond), but it is hard to determine the oscillation parameters. With a fit to the detrended data points (red asterisk) in a form of $F(t) = A \exp(-\frac{t}{\tau}) \sin(\omega t+\phi)$, we obtain the velocity amplitude ($A$), oscillation period ($T$), and the damping time ($\tau$) of the filament. The results show that the oscillation periods at different positions on F1 are exactly the same, i.e., $T = 22$ minutes. This result indicates that the filament was oscillated as a whole harmonic oscillator. It is noted that the largest velocity amplitude is detected at P1, and then it decreases gradually along the filament axis ($A = $\speed{14, 13, 10, and 12} at P1, P2, P3, and P4, respectively), which may indicate the weaken process of the EUV wave which successively converts the wave energy into the mechanical energy of the filament. Due to the lower cadence of the SMART observations, it is hard to determine the exact start time of the oscillation at the four positions. The distance from P1 to P4 is about 150 Mm. If the wave speed is \speed{550}, it needs about 5 min to pass such a distance. However, as one can see in \fig{fig4} (a) -- (d), the oscillation at the four positions show little phase difference, which may indicate that the wave vector is not along the filament axis (see animation 2). We also measure the Doppler velocities of F2 -- F4 with the same method, and the results are shown in \fig{fig4} (f) -- (h). The results show that the oscillation periods (amplitudes) of F2, F3, and F4 are 11 (6), 16 (6), and 16 (10) minutes (\speed{}), respectively. These observational results indicate that different filaments aways oscillate with their own characteristic periods, even though for a common disturbance.

\begin{table*}[thbp]
\begin{center}
\caption{Some parameters of the flare, oscillating filaments, and the jet.\label{tbl1}}
\begin{tabular}{lcccccc}
\tableline\tableline
Items &Flare &F1 &F2  &F3 &F4 &Jet\\
\tableline
Start Time (UT) &22:12:00 &22:24:53 &22:20:53 &22:24:53 &22:26:53  &22:28:46\\
Dist. (Mm)  &$\cdot\cdot\cdot$ &475 &74 &219 &494 &685\\
Time Diff. (s) &$\cdot\cdot\cdot$ &773 &533 &773 &893 &1006\\
Speed (\speed{}) &$\cdot\cdot\cdot$ &648 &138 &298 &556 &680\\
\tableline
\end{tabular}
\tablecomments{In the table, ``Dist.'' is the distance measured from the flare kernel to the filaments and the jet ``Time Diff.'' is the time difference between the start times of the flare and the targets. ``Speed'' is estimated speed of the needed EUV wave, which is obtained by dividing the distance using the time difference.}
\end{center}
\end{table*}

\subsection{The Coronal Jet}
At about 22:28 UT, 16 minutes after the start of the flare, a small jet is observed near the southern boundary of the trans-equatorial coronal hole. According to the time relationship between the jet and the flare, as well as the magnetic field configuration around the jet, we propose that this jet can also be connected to the flare activity in AR11283. The location of the jet is indicated by the arrow in \fig{fig1}(c), while the evolution details are shown in \fig{fig5}.

The magnetic configuration is shown in \fig{fig5}(d), and the boundary of the coronal hole is highlighted with the white dotted curve. Around the jet-base, there is a small magnetic bipole whose positive polarity is close to the coronal hole, in which the magnetic polarity is dominated by negative magnetic field. Such a magnetic configuration is in favor of the occurrence of a jet if the magnetic environment is disturbed. Here, the EUV wave from the northwest (see the white arrow) would push down the open fields in the coronal hole to reconnect with the closed bipole. Before the start of the jet, at the location of the jet-base is a small bright point in both H$\alpha$ and EUV observations (see panels (a) and (g) in \fig{fig5}). At about 22:34 UT, a flare-like brightening is observed at the edge of the bipole, and then a jet originated in-between the bipole and the coronal hole is observed at both H$\alpha$ and EUV wavelengths. This brightening could be interpreted as a direct evidence of the magnetic reconnection occurred between the open fields in the coronal hole and the nearby small closed bipole. The variation of the magnetic flux over the jet-base region is checked, and no significant flux emergence could be found. Therefore, we propose that this jet should be launched by the reconnection between the closed bipole and the ambient open fields in the coronal hole, which were pushed down to reconnect with the bipole and thus resulted in the jet. It is noted that this jet did not show obvious rotating motion as it has been reported in previous studies \citep[e.g.,][]{shen11a,shen12e,chen12,hong13}. This may indicate the different magnetic field environments and driving mechanisms of the present jet and the rotating jets. For rotating jets, the bipole is usually strongly twisted before eruption, and the releasing of the stored twist can cause the rotation motion \citep{shen11a}.

\subsection{The Relationship between the EUV Waves and Oscillating Filaments}
So far, we have described the solar activities in association with the X2.1 flare on September 06, including the EUV waves, the oscillating filaments, and the jet. Here, we try to establish the relationship between the EUV waves and other activities. We first measure the distance from the flare kernel to the filaments and the jet, and then estimate the required wave speed to trigger the filament oscillations and the jet. Here, we assume that the wave started at the same time with the flare. The parameters of the flare, the winking filaments, and the jet are listed in \tbl{tbl1}. In the table, the start time of the filament oscillations and the jet are determined from SMART H$\alpha$ observations. By checking the start time of the these activities, we find that the start times of the filament oscillations and the jet appear linearly depending on their distance to the flare kernel. We further calculate the required speed of a possible wave to trigger the oscillation of the filaments and the jet. We find that to launch the oscillation of F1, the required speed is about \speed{648}, while that is about \speed{680} for the jet. These results suggest that the fast wave, whose peed is \speed{554} along S2 (see \fig{fig2} (b)), could be the trigger of the oscillation of F1 and the jet. The required speeds to launch the oscillation of F2, F3, and F4 are about \speed{138, 298, and 556}, respectively. These speeds are too small by comparison with the measured speed of the fast EUV wave (\speed{850}). However it seems close to the speed of the slow EUV wave (\speed{338}). In spite of this, we still can not say that the oscillation of F2 -- F4 are launched by the slow EUV wave, because the appearance location of the slow wave was between F3 and F4. In fact, due to the large cadence of the SMART data (2 min), it can result in a large error when we determine the start time of the filament oscillations. In addition, as indicated in previous studies, the start time of an EUV wave is often posterior to the start of the associated flare several minutes, and the generation location is also not the flare kernel. Since these reasons and the short distance from the flare kernel to F2 and F3, it is difficult to figure out the relationship between the fast EUV wave and F2 -- F3. However, since the start of the oscillations of F2 -- F4, as seen in the H$\alpha$ observations, are in a proper time order, we still propose that the launch of the oscillation of the three filaments are caused by the fast EUV wave.

\section{Conclusions \& Discussions}
In this paper, we present an observational study of a chain of winking (oscillation) filaments and a small jet that occurred in a proper time order but separated far away from each other. The event initiated from an X2.1 flare in AR11283, after which  a remarkable EUV wave is observed mainly in the northwest direction, which swept over three filaments in the west hemisphere and launched there oscillations. It is interested that oscillation of a long filament and a small jet in the east hemisphere are also observed following the flare. With the excellent observations taken by the SMART and the {\it SDO} instruments, we find that all these successive solar activities, including four oscillating filaments and a jet, were dynamically connected to the global EUV wave. The main analysis results could be summarized as follows.
\begin{enumerate}
\item A remarkable EUV wave is observed a few minutes after the start of the X2.1 flare, which propagates outward mainly in the northwest side of AR11283 and with an average projection speed (deceleration) of about \speed{850} (\acc{-224}). Behind the EUV wave, we detect another slow wave whose speed is about \speed{338}. In the east of AR11283, the wave signature is very weak or even invisible in the imaging observations, and the average speed is about \speed{554}. Based on our analysis, we propose that the shape of the EUV wave should be like a dome as what has been reported in \cite{vero10}. In this line of thought, we propose that the prominent wave structure in the northwest of AR11283 should be the upward expanding dome structure, while the weak wave signature in the east hemisphere should be the surface EUV wave that is driven by the associated CME at the initial stage and then propagates freely on the surface. In addition, we note that there is no detectable chromosphere Moreton wave associated with the EUV wave, which may imply that the downward compression to the dense chromosphere of the fast EUV wave is too weak, maybe it is possibly due to a large angle of the EUV wave to the solar surface \citep{tama13}.
\item In the SMART H$\alpha$ line-wing observations, four successive winking (oscillating) filaments are observed, and the start of their oscillations are in a proper time order depending on their distances to the flare kernel. All the filaments oscillated for about two or three cycles and then regained their equilibriums instead of eruptions. The intensity fluctuation amplitudes in the H$\alpha$ line-center and the EUV observations are very subtle, which suggests a relative small oscillation amplitudes of the filaments. By analyzing the temporal and spatial relationships between the oscillating filaments and the EUV wave, we propose that all the filament oscillations are triggered by one common agent, i.e., the weak (or invisible) surface EUV wave. It is also noted that some filaments did not disturbed during the event, no matter how close they are to the flare kernel (for example, F6, see \fig{fig1} (a)). Such phenomenon is also noted by \cite{okam04}, and the detailed reasons need further investigations.
\item Using the Beckers' cloud model, Some physical parameters of the oscillating filaments are derived. The results show that the oscillation periods of the filaments range from 11 to 22 minutes, the Doppler velocity amplitudes are from \speed{6 to 14}, and the damping times are from 25 to 60 minutes. These results may indicate inherent property of the filaments, namely, a  filament always oscillates with its own characteristic parameters. It is noted that the filament oscillation launched by large-scale EUV wave often has a large amplitude at the very beginning, but it will quickly reduce to a moderate amplitude and then undergoes an ordinary damping oscillation process. This characteristic could be recognized as an observational characteristic of filament oscillations that are trigged by pulse-like EUV waves.
\item A small jet is observed close to the southern boundary of a large trans-equatorial coronal hole following the sequence of the successive filament oscillations. Considering the magnetic configuration around the jet-base and the ejection details of the jet, we propose that this jet should be caused by the magnetic reconnection between the biple and the ambient open fields in the coronal hole, which were pushed down by the EUV wave and thereby trigger the reconnection.
\item Based on the analyzing of the temporal and spatial relationships among the EUV waves and the oscillating filaments, as well as the small jet. We propose that all the observed filament oscillations and the jet are triggered by the large-scale surface EUV wave, which could be the physical linkage of many successive but separated solar activities. Even though the EUV wave is very weak or even invisible in coronal imaging observations, it also can trigger the occurrence of many different solar activities. The oscillation of F1 and the occurrence of the jet in the present case could be a good example for such a conclusion. Vice versa, the observation of winking (oscillation) filaments can provide an effective way to diagnose the arriving and the property of weak or invisible waves in the solar atmosphere.
\end{enumerate}

Many studies have indicated that EUV waves often have their chromospheric counterparts, i.e., Moreton waves. It seems that how deep a coronal wave can reach down to the lower dense atmosphere depends on its strength and the angle between the wave and the solar surface. For example, in a powerful X6.9 flare event, \cite{shen12c} found that the EUV wave can reach down to the upper photosphere or the bottom of the chromosphere. In the present case, however, no chromosphere Moreton wave can be detected in H$\alpha$ observations, suggesting that the observed fast EUV should be a weak one (at least for the surface wave), or it has a large angle relevant to the solar surface \citep{tama13}.

A slow wave behind the fas EUV wave was often interpreted as a pseudo-wave in previous studies \citep[e.g.,][]{chen02,shen12c}. So far there are several interpretations for such an EUV wave, which are related to a current shell or successive restructuring of field lines caused by the associated CME \citep[e.g.,][]{chen02,dela07,attr07}. In addition, some authors claimed that both the wave and non-wave models are required to explain the EUV waves \citep[e.g.,][]{cohe09,liu10,down11}. Recently, \cite{gosa12} found that a filament is triggered by both the fast and slow EUV waves successively in a flaring event, in which the slow EUV wave resulted in a phase change of the filament oscillation. Their observation seems supporting the scenarios that such a slow EUV wave should be a real MHD wave. In the present case, the observed slow EUV wave is possibly a secondary wave triggered by the fast EUV wave when it interacts with other magnetic structures on the path. On the other hand, since the northwestward wave structure is considered as the upward dome structure of the EUV wave, there is another possible that the slow wave may represent the surface wave launched by the associated CME. In this sense, one can interpret the slow EUV wave observed in the present case as a real MHD wave in the physical nature.

For the oscillating filaments, the observed parameters could be applied into filament seismology to estimate the other important filament parameters that are difficult to obtain with direct measuements. With the method proposed by \cite{hyde66} and the measured oscillation parameters, one can derive the radial component of the filament magnetic field which supports the filament mass. In Hyder's method, the relation between the radial magnetic field and the oscillation period and damping time can be written as $B_{\rm r}^{2} = \pi \rho \, r_{\rm 0}^{2} \, [4 \, \pi ^{2} \, (\frac{1}{T})^{2} + (\frac{1}{\tau})^{2}]$, where $B_{\rm r}$ is the radial magnetic component, $\rho$ is the density of the filament mass, $r_{\rm 0}$ is the scale height of the filament, $T$ is the filament oscillation period, and $\tau$ is the damping time. If we use the value $\rho = 4 \times 10^{-14} \, {\rm gm/cm^{3}}$, i.e., $n_{\rm e} = 2 \times 10^{10} \, {\rm cm^{-3}}$, the above equation can be rewritten as the form $B_{\rm r}^{2} = 4.8 \times 10^{-12} \, r_{0}^{2} \, [(\frac{1}{T})^{2} + 0.025 \, (\frac{1}{\tau})^{2}]$. With the measured oscillation periods and damping times, we obtain that the value of the radial component of the filaments' magnetic fields $B_{\rm r}$ are 5.0, 10.0, 7.0, and 7.0 Gauss for the filaments F1 -- F4, respectively. In the calculation, we use the value $r_{\rm 0} = 3 \times 10^{9}$ cm, the same with \cite{hyde66}. The estimated filament magnetic fields are in agreement with the values obtained by direct measuring or inferred from the analysis of polarization of filament lines \citep[e.g.,][]{ziri61,warw65,lee65}.

In addition, one can also estimate the total maximum kinetic energy of the oscillating filaments by using the measured maximum Doppler velocities. The maximum kinetic energy of an oscillating filament is equal to the work done by the pressure gradient force in the wave as it accelerates the filament. Here, we take the values measured at P1 on F1 as an example. The maximum redshift velocity at P1 is \speed{21}, and therefore, the maximum kinetic energy is $E = \frac{1}{2} \, m v^{2} \approx 9.0 \times 10^{19} \, {\rm J}$, where $m$ is the mass of the filament and we assume it as $4 \times 10^{14} \, {\rm g}$ \citep{gilb06}. The obtained value of the oscillating filament is in agreement with the predicted energy required to induce the oscillations of a quiescent filament, namely, $\sim 10^{19}$ -- $10^{20} \,  {\rm J}$ \citep{klec69}. The maximum kinetic energy of the oscillating filament is comparable with the value of the minimum energy of EUV waves, which is of the order of $10^{19} \, {\rm J}$ \citep{ball05}. These results indicate that the energy of the EUV wave is sufficient to launch filament oscillations on the path.

The event presented in this paper is a good example of the so-called sympathetic solar eruptions, i.e., a number of successive but separated solar eruptive activities occurring within a short timescale. The key question of sympathetic eruptions is whether the close temporal correlation between successive solar activities is purely coincidental or causally linked. Therefore, the searching for the physical linkage among the successive solar activities are important for interpreting sympathetic eruptions. Many studies have shown that most of sympathetic eruptions are dynamically connected, and the basic mechanism is often of a magnetic nature. For example, \cite{schr11} found that successive events could be connected by a system of separatrices, separators, and quasi-separatrix layers. \cite{shen12d} found that magnetic reconnection is important for the production of sympathetic filament eruptions. In addition, these studies also indicate the importance of the magnetic topological structure of the global coronal field in the production of sympathetic eruptions \citep[see also,][]{jian08,toro11,tito12,lync13,schr13,kong13}. In the present case, our analysis results indicate that EUV waves can be a good agent for connecting successive but separated solar activities.

\acknowledgments {\em SDO} is a mission for NASA's Living With a Star (LWS) Program. The authors would like to thank the excellent H$\alpha$ data provided by the SMART team at Hida Observatory, Japan. The authors would thank an anonymous referee for many valuable comments and suggestions for improving the quality of this paper. Y.S. is supported by the research fellowship of Kyoto University, and he would like to thank the staff of Kwasan and Hida Observatories for their support and helpful comments. This work is supported by the Open Research Programs of the Key laboratory of Dark Matter and Space Astronomy of CAS (DMS2012KT008), and the CAS Western Light Youth Project.


\begin{thebibliography}{}
\bibitem[Arregui et al.(2012)]{arre12}
Arregui, I., Oliver, R., \& Ballester, J. L. 2012, Living Rev. Sol. Phys., 9, 2
\bibitem[Asai et al.(2012)]{asai12}
Asai, A., Ishii, T. T., Isobe, H., et al. 2012, \apjl, 745, L18
\bibitem[Athay \& Moreton(1961)]{atha61}
Athay, R. G., \& Moreton, G. E. 1961, \apj, 133, 935
\bibitem[Attrill et al.(2007)]{attr07}
Attrill, G. D. R.,Harra, L. K., vanDriel-Gesztelyi, L., \& D\'{e}moulin, P. 2007, \apj, 656, L101
\bibitem[Ballai et al.(2005)]{ball05}
Ballai, I., Erd\'{e}lyi, R., \& Pint\'{e}r, B. 2005, \apjl, 633, L145
\bibitem[Beckers (1964)]{beck64}
Beckers, J. M. 1964, PhD Thesis, University of Utrecht
\bibitem[Bi et al.(2012)]{bi12}
Bi, Y., Jiang, Y., Li, H., Hong, J., \& Zheng, R. 2012, \apj, 758, 42
\bibitem[Bi et al.(2013)]{bi13a}
Bi, Y., Jiang, Y., Yang, J., et al. 2013, \apj, 773, 162
\bibitem[Chen et al.(2013)]{chen13}
Chen, H. D., Ma, S. L., \& Zhang, J. 2013, \apj, 778, 70
\bibitem[Chen et al.(2012)]{chen12}
Chen, H. D., Zhang, J., \& Ma, S. L. 2012, Res. Astron. Astrophys., 12, 573
\bibitem[Chen et al.(2008)]{chen08}
Chen, P. F., Innes, D. E., \& Solanki, S. K. 2008, \aap, 484, 487
\bibitem[Chen et al.(2011)]{chen11}
Chen, P. F., \& Wu, Y. 2011, \apjl, 732, L20
\bibitem[Chen et al.(2002)]{chen02}
Chen, P. F., Wu, S. T., Shibata, K., \& Fang, C. 2002, \apjl, 572, L99
\bibitem[Cohen et al.(2009)]{cohe09}
Cohen, O., Attrill, G. D. R., Manchester, W. B., IV, \& Wills-Davey, M. J. 2009, \apj, 705, 587
\bibitem[Dai et al.(2013)]{dai13}
Dai, Y., Ding, M. D., \& Guo, Y. 2013, \apjl, 773, L21
\bibitem[Delann\'{e}e et al.(2007)]{dela07}
Delann\'{e}e, C. Hochedez, J. F., \& Aulanier, G. 2007, \aap, 465, 603
\bibitem[Dodson(1949)]{dods49}
Dodson, H. W. 1949, \apj, 110, 382
\bibitem[Downs et al.(2011)]{down11}
Downs, C., Roussev, I. I., Van der Holst, B., et al. 2011, \apj, 728, 2
\bibitem[Dyson(1930)]{dyso30}
Dyson, F. 1930, \mnras, 91, 239
\bibitem[Eto et al.(2002)]{eto02}
Eto, S., Isobe, H., Narukage, N., et al. 2002, \pasj, 54, 481
\bibitem[Feng et al.(2013)]{feng13}
Feng, L., Wiegelmann, T., Su, Y., et al. 2013, \apj, 765, 37
\bibitem[Foullon et al.(2004)]{foul04}
Foullon, C., Verwichte, E., \& Nakariakov, V. M. 2004, \aap, 427, L5
\bibitem[Foullon et al.(2009)]{foul09}
Foullon, C., Verwichte, E., \& Nakariakov, V. M. 2009, \apj, 700, 1658
\bibitem[Gilbert et al.(2008)]{gilb08}
Gilbert, H. R., Daou, A. G., Young, D., Tripathi, D., \& Alexander, D. 2008, \apj, 685, 629
\bibitem[Gilbert et al.(2006)]{gilb06}
Gilbert, H. R., Falco, L. E., Holzer, T. E., \& MacQueen, R. M. 2006, \apj, 641, 606
\bibitem[Gosain \& Foullon(2012)]{gosa12}
Gosain, S., \& Foullon, C. 2012, \apj, 761, 103
\bibitem[Hershaw et al.(2011)]{hers11}
Hershaw, J., Foullon, C., Nakariakov, V. M., \& Verwichte, E. 2011, \aap, 531, A53
\bibitem[Hillier et al.(2013)]{hill13}
Hillier, A., Morton, R. J., \& Erd\'{e}lyi, R. 2013, \apjl, 779, L16
\bibitem[Hong et al.(2013)]{hong13}
Hong, J. C., Jiang, Y. C., Yang, J. Y., et al. 2013, Res. Astron. Astrophys., 13, 253
\bibitem[Hyder(1965)]{hyde65}
Hyder, C. L. 1965, \apj, 141, 1374
\bibitem[Hyder(1966)]{hyde66}
Hyder, C. L. 1966, Z. Astrophys. 63, 78
\bibitem[Ishii et al.(2013)]{ishi13}
Ishii, T. T., Kwawate, T., Nakatani, Y., et al. 2013, \pasj, 65, 39
\bibitem[Isobe \& Tripathi(2006)]{isob06}
Isobe, H., \& Tripathi, D. 2006, \aap, 449, L17
\bibitem[Isobe et al.(2007)]{isob07}
Isobe, H., Tripathi, D., Asai, A., \& Jain, R. 2007, \solphys, 246, 89
\bibitem[Jackiewicz \& Balasubramaniam(2013)]{jack13}
Jackiewicz, J., \& Balasubramaniam, K. S. 2013, \apj, 765, 15
\bibitem[Jiang et al.(2013)]{jian13}
Jiang, C., Feng, X., Wu, S. T., \& Hu, Q. 2013, \apjl, 771, L30
\bibitem[Jiang et al.(2014a)]{jian14a}
Jiang, C., Wu, S. T., Feng, X., \& Hu, Q. 2014a, \apj, 780, 55
\bibitem[Jiang et al.(2014b)]{jian14b}
Jiang, C., Wu, S. T., Feng, X., \& Hu, Q. 2014b, \apj, submitted
\bibitem[Jiang et al.(2008)]{jian08}
Jiang, Y. C., Shen, Y. D., Yi, B., Yang, J. Y., \& Wang, J. X. 2008, \apj, 667, 669
\bibitem[Jing et al.(2003)]{jing03}
Jing, J., Lee, J., Spirock, T. J., et al. 2003, \apjl, 584, L103
\bibitem[Jing et al.(2006)]{jing06}
Jing, J., Lee, J., Spirock, T. J., \& Wang, H. 2006, \solphys, 236, 97
\bibitem[Kippenhahn \& Schl\"{u}ter(1957)]{kipp57}
Kippenhahn, R., \& Schl\"{u}ter, A. 1957, ZAp, 43, 36
\bibitem[Kleczek \& Kuperus(1969)]{klec69}
Kleczek, J., \& Kuperus, M. 1969, \solphys, 6, 72
\bibitem[Kong et al.(2013)]{kong13}
Kong, D. F., Yan, X. L., \& Xue, Z. K. 2013, \apss, 348, 303
\bibitem[Kuperus \& Raadu(1974)]{kupe74}
Kuperus, M., \& Raadu, M. A. 1974, \aap, 31, 189
\bibitem[Kurokawa et al.(1995)]{kuro95}
Kurokawa, H., Ishiura, K., Kimura, G., et al. 1995, J. Geomag. Geoelectr., 47, 1043
\bibitem[Lee et al.(1965)]{lee65}
Lee, R. H., Rust, D. M., \& Zirin, H. 1965, Appl. Opt. 4, 1081
\bibitem[Lemen et al.(2012)]{leme12}
Lemen, J. R., Title, A. M., Akin, D. J., et al. \solphys, 275, 17
\bibitem[Li \& Zhang(2012)]{li12}
Li, T., \& Zhang, J. 2012, \apjl, 760, L10
\bibitem[Liu et al.(2013)]{liu13}
Liu, R., Liu, C., Xu, Y., et al. 2013, \apj, 773, 166
\bibitem[Liu et al.(2010)]{liu10}
Liu, W., Nitta, N. V., Schrijver, C. J., Title, A. M., \& Tarbell, T. D. 2010, \apj, 723, L53
\bibitem[Liu et al.(2012)]{liu12}
Liu, W., Ofman, L., Nitta, N. V., et al. 2012, \apj, 753, 52
\bibitem[Liu et al.(2011)]{liu11}
Liu, W., Title, A. M., Zhao, J. W., et al. 2011, \apjl, 736, 13L
\bibitem[Luna et al.(2012)]{luna12a}
Luna, M., D\'{i}az, A. J., \& Karpen, J. 2012, \apj, 757, 98
\bibitem[Luna \& Karpen(2012)]{luna12b}
Luna, M., \& Karpen, J. 2012, \apjl, 750, L1
\bibitem[Lynch \& Edmondson(2013)]{lync13}
Lynch, B. J., \& Edmondson, J. K. 2013, \apj, 764, 87
\bibitem[Moreton \& Ramsey(1960)]{more60}
Moreton, G. E., \& Ramsey, H. E. 1960, \pasp, 72, 357
\bibitem[Morimoto \& Kurokawa(2003)]{mori03}
Morimoto, T., \& Kurokawa, H. 2003, \pasj, 55, 503
\bibitem[Morimoto et al.(2010)]{mori10}
Morimoto, T., Kurokawa, H., Shibata, K., \& Ishii, T. T. 2010, \pasj, 62, 939
\bibitem[Nagata et al.(2014)]{naga14}
Nagata, S., et al. 2014, \pasj, to be published
\bibitem[Newton(1935)]{newt35}
Newton, H. W. 1935, \mnras, 95, 650
\bibitem[Nitta et al.(2013)]{nitt13}
Nitta, N. V., Schrijver, C. J., Title, A. M., \& Liu, W. 2013, \apj, 776, 58
\bibitem[Okamoto et al.(2004)]{okam04}
Okamoto, T. J., Nakai, H., Keiyama, A., et al. 2004, \apj, 608, 1124
\bibitem[Oliver \& Ballester(2002)]{oliv02}
Oliver, R., \& Ballester, J. L. 2002, \solphys, 206, 45
\bibitem[Pesnell et al.(2012)]{pesn12}
Pesnell, W. D., Thompson, B. J., \& Chamberlin, P. C. 2012, \solphys, 275, 3
\bibitem[Pint\'{e}r et al.(2008)]{pint08}
Pint\'{e}r, B., Jain, R., Isobe, H. 2008, \apj, 680, 1560
\bibitem[Ramsey \& Smith(1966)]{rams66}
Ramsey, H. E., \& Smith, S. F. 1966, \aj, 71, 197
\bibitem[Romano \& Zuccarello(2013)]{roma13}
Romano, P., \& Zuccarello, F. 2013, Mem. S.A.It., 84, 363
\bibitem[Ruan et al.(2014)]{ruan14}
Ruan, G., Chen, Y., Wang, S., et al. 2014, 784, 165
\bibitem[Schou et al.(2012)]{scho12}
Schou, J., Borrero, J. M., Norton, A. A., et al., \solphys, 275, 327
\bibitem[Schrijver \& Title (2011)]{schr11}
Schrijver, C. J., \& Title, A. M. 2011, \jgr, 116, A04108
\bibitem[Schrijver et al.(2013)]{schr13}
Schrijver, C. J., Title, A. M., Yeates, A. R., \& DeRosa, M. L. 2013, \apj, 773, 93
\bibitem[Shen \& Liu(2012a)]{shen12a}
Shen, Y. D., \& Liu, Y. 2012a, \apj, 753, 53
\bibitem[Shen \& Liu(2012b)]{shen12b}
Shen, Y. D., \& Liu, Y. 2012b, \apj, 754, 7.
\bibitem[Shen \& Liu(2012c)]{shen12c}
Shen, Y. D., \& Liu, Y. 2012c, \apjl, 752, L23.
\bibitem[Shen et al.(2011b)]{shen11b}
Shen, Y. D., Liu, Y., \& Liu, R. 2011b, Res. Astron. Astrophys., 11, 594
\bibitem[Shen et al.(2012d)]{shen12d}
Shen, Y. D., Liu, Y., \& Su, J. T. 2012d, \apj, 750, 12
\bibitem[Shen et al.(2012e)]{shen12e}
Shen, Y. D., Liu, Y., Su, J. T., \& Deng, Y. Y. 2012e, \apj, 754, 164
\bibitem[Shen et al.(2011a)]{shen11a}
Shen, Y. D., Liu, Y., Su, J. T., \& Ibrahim, A. 2011a, \apjl, 735, L43
\bibitem[Shen et al.(2013a)]{shen13a}
Shen, Y. D., Liu, Y., Su, J. T., et al. 2013a, \apj, 773, L33.
\bibitem[Shen et al.(2013b)]{shen13b}
Shen, Y. D., Liu, Y., Su, J. T., et al. 2013b, \solphys, 288, 585
\bibitem[Tamazawa et al.(2013)]{tama13}
Tamazawa, H., et al. 2013, report in the Solar Seminar of Kwasan Observatory
\bibitem[Tandberg-Hanssen(1995)]{tand95}
Tandberg-Hanssen, E. 1995, The Nature of Solar Prominences, (Dordrecht: Kluwer)
\bibitem[Titov et al.(2012)]{tito12}
Titov, V. S., Mikic, Z., T\"{o}r\"{o}k, T., Linker, J. A., \& Panasenco, O. 2012, \apj, 759, 70
\bibitem[T\"{o}r\"{o}k et al.(2011)]{toro11}
T\"{o}r\"{o}k, T., Panasenco, O., Titov, V. S., et al. 2011, \apjl, 739, L63
\bibitem[Tripathi et al.(2009)]{trip09}
Tripathi, D., Isobe, H., \& Jain, R. 2009, \ssr, 149, 283
\bibitem[Tziotziou(2007)]{tzio07}
Tziotziou, K. 2007, in The Physics of Chromospheric Plasmas, eds. P. Heinzel, I. Dorotovi\v{c}, \& R. J. Rutten, ASP Conf. Ser., 368, 217
\bibitem[Ueno et al.(2004)]{ueno04}
Ueno, S., Nagata, S., Kitai, R., \& Kurokawa, H. 2004, in ASP Conf. Ser. 325, 319
\bibitem[Uchida(1968)]{uchi68}
Uchida, Y. 1968, \solphys, 4, 30
\bibitem[Veronig et al.(2010)]{vero10}
Veronig, A. M., Muhr, N., Kienreich, I. W., Temmer, M., \& Vrv{s}nak, B. 2010, \apjl, 716, L57
\bibitem[Verwichte et al.(2013)]{verw13}
Verwichte, E., Van Doorsselaere, T., Foullon, C., \& White, R. S. 2013, \apj, 767, 16
\bibitem[Vr\v{s}nak et al.(2007)]{vrsn07}
Vr\v{s}nak, B., Veronig, A. M., Thalmann, J. K., \& \v{Z}ic, T. 2007, \aap, 471, 295
\bibitem[Warwick \& Hyder(1965)]{warw65}
Warwick, J. W., \& Hyder, C. L. 1965, \apj, 141, 1362
\bibitem[Woods et al.(2012)]{wood12}
Woods, T. N., Eparvier, F. G., \& Hock, R. et al. 2012, \solphys, 275, 115
\bibitem[Xue et al.(2013)]{xue13}
Xue, Z. K., Qu, Z. Q., Yan, X. L., Zhao, L., \& Ma, L. 2013, \aap, 556, A152
\bibitem[Yan et al.(2012a)]{yan12a}
Yan, X. L., Qu, Z. Q., \& Kong, D. F. 2012a, \aj, 143, 56
\bibitem[Yan et al.(2012b)]{yan12b}
Yan, X. L., Qu, Z. Q., Kong, D. F., \& Xu, C. L. 2012b, \apj, 754, 16
\bibitem[Zirin \& Severny(1961)]{ziri61}
Zirin, H., \& Severny, A. B. 1961, Observatory, 81, 155
\bibitem[Zirker et al.(1998)]{zirk98}
Zirker, J. B., Engvoid, O., \& Martin, S. F. 1998, \nat, 396, 440
\bibitem[Zhang et al.(2012)]{zhan12}
Zhang, Q. M., Chen, P. F., Xia, C., \& Keppens, R. 2012, \aap, 542, A52
\bibitem[Zhang et al.(2013)]{zhan13}
Zhang, Q. M., Chen, P. F., Xia, C., Keppens, R., \& Ji, H. S. 2013, \aap, 554, A124
\end{thebibliography}
\end{document}